# Enhanced thermal conductivity of (010) $(Al_xGa_{1-x})_2O_3$ epitaxial films utilizing indium-catalyzed molecular beam epitaxy

Shivashree Gowda[1], Stephen Schaefer[2], Ethan A. Scott[1], Samreen Khan[1], Patrick E. Hopkins[1,3,4,a), and M. Brooks Tellekamp[2,a)]

[1]Department of Mechanical and Aerospace Engineering, University of Virginia, Charlottesville, VA 22904, USA

[2]National Laboratory of the Rockies, Golden, CO 80401, USA

[3]Department of Material Science and Engineering, University of Virginia, Charlottesville, VA 22904, USA

[4]Department of physics, University of Virginia, Charlottesville, VA 22904, USA

[a)] **Authors to whom correspondence should be addressed**:

brooks.tellekamp@nlr.gov, peh4v@virginia.edu

## Abstract

$(Al_xGa_{1-x})_2O_3/\beta\text{-}Ga_2O_3$ transistors are an emerging candidate for high-power and high-frequency electronic devices. $\beta\text{-}Ga_2O_3$ in particular is notably limited for anisotropic low thermal conductivity, which is further reduced in $(Al_xGa_{1-x})_2O_3$ due to alloy and other defect-driven phonon scattering mechanisms. In this work we show that the thermal conductivity of (010) oriented $(Al_xGa_{1-x})_2O_3$ thin films for $0.01 < x < 0.20$, measured using time-domain thermoreflectance (TDTR), is limited by alloy scattering without observable adverse scattering by additional defects. This is enabled through the synthesis of the $(Al_xGa_{1-x})_2O_3$ films using molecular beam epitaxy (MBE) on $\beta\text{-}Ga_2O_3$ substrates, leveraging indium-catalyzed growth to suppress dislocation formation and phase separation to achieve single-phase pseudomorphic films up to x = 0.2. This growth process improved the thermal conductivity of $(Al_xGa_{1-x})_2O_3$ by 2X as compared to previously reported values. For increasing Al composition (x), we observe a steady decline in $(Al_xGa_{1-x})_2O_3$ thermal conductivity due to alloy scattering which is validated using virtual crystal approximation (VCA) model. We also show that the thermal boundary conductance across the $Al/(Al_xGa_{1-x})_2O_3$ interface is reduced with increasing x, which we posit is due to the stiffening of the $(Al_xGa_{1-x})_2O_3$ acoustic modes with increasing x by comparing experimental results with a diffuse mismatch model (DMM). Overall, these thermal characteristics provide valuable insights for designing heterostructures with optimized interfaces and composition.

Alloying is an essential process for the design of semiconductor devices. The ability to tune properties between end-members enables spatial control of charge carrier populations, electric fields, and the overall energy landscape. This strategy has enabled solid state optoelectronics such as solar cells, photodetectors, and light emitting diodes by controlling where and when electrons and holes interact in the solid state, and how efficiently they are injected into or extracted from a device. Control of energy landscapes in high-frequency and high-power electronic devices is equally important, where alloys are used in modulation-doped field-effect transistors (MODFETs) or high electron mobility transistors (HEMTs) to confine carriers in a 2-dimensional electron gas (2DEG). Alloying with wider bandgap materials is also used to passivate crystalline surfaces with high quality dielectrics to prevent leakage or premature breakdown.

While alloying is critical for these high frequency and power devices, it also presents a challenge – alloy scattering reduces thermal conductivity which negatively impacts the performance of electronic devices. This is a known design constraint in MODFETs and HEMTs, where high current densities and localized heating result in current collapse, gate leakge, and catastrophic failure in multi-device modules. It is therefore critical to accurately measure alloy-scattering limited thermal conductivity, in the low dislocation limit, of ultra-wide band gap (UWBG) alloys to understand their performance limits and accurately design thermal management systems for next generation power electronics.

Monoclinic gallium oxide ($\beta$-$Ga_2O_3$) has emerged as a promising UWBG semiconductor for next-generation high-power and high-frequency electronics.[1,2] Its favorable intrinsic properties include a wide bandgap of 4.6–4.8 eV,[3,4] a theoretical breakdown field of 8 MV/cm,[5,6] efficient n-type doping, and an electron saturation velocity of $2 \times 10^7$ cm/s.[7] Crucially, native bulk $\beta$-$Ga_2O_3$ single crystals can be synthesized via cost-effective melt-growth techniques, enabling scalable substrate manufacturing.[8,9]

Despite these advantages, the performance of $\beta$-$Ga_2O_3$ device is constrained by low thermal conductivity and bulk electron mobility.[10–13] To mitigate transport limitations, $(Al_xGa_{1-x})_2O_3$/$\beta$-$Ga_2O_3$ MODFETs are utilized.[14–16] At the $(Al_xGa_{1-x})_2O_3$/$\beta$-$Ga_2O_3$ heterostructure interfaces, confinement from conduction band offset and deliberate spatial separation of donors from the channel results in a two-dimensional electron gas (2DEG).[17,18] This 2DEG channel achieves high mobility and reduced sheet resistance by isolating carrier transport from ionized impurity scattering, significantly outperforming conventional metal-oxide-semiconductor field-effect transistor (MOSFET) architectures in high-frequency applications.[19–22]

$\beta$-$Ga_2O_3$ can be effectively alloyed with $Al_2O_3$ at low compositions for band structure engineering, however, $\beta$-$Ga_2O_3$ and $Al_2O_3$ have different thermodynamic ground state structures, and excess Al in the alloy results in degraded crystal quality and phase-separation.[23–25] The Al mole-fraction at which phase separation occurs depends on the $\beta$-$Ga_2O_3$ crystal orientation; for (010) orientation, phase separation was seen to occur above x=0.2 in $(Al_xGa_{1-x})_2O_3$.[26] This is critical for MODFETs because the sheet charge density which reduces channel resistance is proportional to the conduction band offset. Increasing Al composition, if it could be accomplished without crystal quality degradation or phase separation, would lead to higher performance transistors.

At the $(Al_xGa_{1-x})_2O_3/\beta$-$Ga_2O_3$ heterostructure interface, high current densities within the sub-10 nm 2DEG channel induce localized hot-spot formation due to inelastic electron-phonon scattering.[27] For effective heat dissipation, the thermal conductivity of $(Al_xGa_{1-x})_2O_3$ should be maximized which has been to date limited due to extrinsic factors such as dislocations and phase separation. Prior investigations of (-201) oriented $(Al_xGa_{1-x})_2O_3$ thin films grown heteroepitaxially on c-plane sapphire[28] and homoepitaxially on (-201) $\beta$-$Ga_2O_3$ substrates[29] reported a nearly composition-independent thermal conductivity of ~4 ± 1 W $m^{-1}$ $K^{-1}$ for Al compositions up to 50%. In a recent study, (010) oriented $(Al_{0.18}Ga_{0.82})_2O_3$ films measured using frequency-domain thermoreflectance (FDTR) yielded a thermal conductivity of 3.6 W $m^{-1}$ $K^{-1}$.[30] Integrating this parameter into technology computer-aided design (TCAD) simulations revealed a 12% increase in cross-plane MODFET temperature rise for a gate-to-source voltage of 5V relative to $Ga_2O_3$ MOSFETs.[30] This temperature rise further intensified when (010) was replaced with (-201) oriented films.[30] These results emphasize that optimizing the thermal conductivity of (010) oriented $(Al_xGa_{1-x})_2O_3$ alloys is more advantageous compared to (-201) orientated film to improve thermal management in high-power MODFETs.

In this work, we investigate the thermal conductivity of (010) oriented $(Al_xGa_{1-x})_2O_3$ thin films using time-domain thermoreflectance (TDTR) across Al compositions ranging from x = 0.011 to 0.206. These films are synthesized by molecular beam epitaxy (MBE) on Fe-doped $\beta$-$Ga_2O_3$ substrates. Indium-catalyzed growth is used to suppress phase separation and extended defect formation for Al fractions up to x = 0.206, dramatically improving overall crystalline quality compared to films grown without an indium catalyst. Using time-domain thermoreflectance (TDTR), we demonstrate that control of phase separation and crystal quality yields the highest thermal conductivity values reported to date for $(Al_xGa_{1-x})_2O_3$ alloys at composition x = 0.137 and 0.206. We also demonstrate that thermal conductivity in these films is limited by intrinsic alloy scattering through a monotonic reduction in thermal conductivity with increasing Al composition, validated using virtual crystal approximation (VCA) calculations. Furthermore, we measure the thermal boundary conductance (TBC) at the $Al/(Al_xGa_{1-x})_2O_3$ interface, where thin Al is used as a transducer, to resolve composition-dependent phonon scattering dynamics, showing a monotonic reduction in TBC that follows diffuse mismatch model (DMM) predictions,[31] suggesting that stiffening of the $(Al_xGa_{1-x})_2O_3$ lattice with increased Al composition reduces phonon TBC. Ultimately, these results can be used for more accurate design of $(Al_xGa_{1-x})_2O_3$/$\beta$-$Ga_2O_3$ heterostructure field-effect transistors (HFETs) with better thermal management.

$(Al,Ga)_2O_3$ thin films were grown on (010) Fe:doped $\beta$-$Ga_2O_3$ substrates (Novel Crystal Technology) using plasma-assisted molecular beam epitaxy (MBE). Indium was used as a catalyst to increase growth rate at high temperatures which promote improved surface morphology and crystal quality without incorporating indium into the film.[32] Further details of the growth conditions are reported in the Supporting Information.

High resolution X-ray diffraction (XRD) was performed using both a Rigaku Smartlab and a Panalytical MRD Pro with 2-bounce or 4-bounce monochromated Cu-Kα X-rays. Rocking curves were taken by rocking the ω-axis with narrow diffracted beam slits, < 0.5 mm, to measure mosaicity of the alloys. Reciprocal space maps (RSMs) were taken in asymmetric geometry around the (420) reflection to analyze strain.

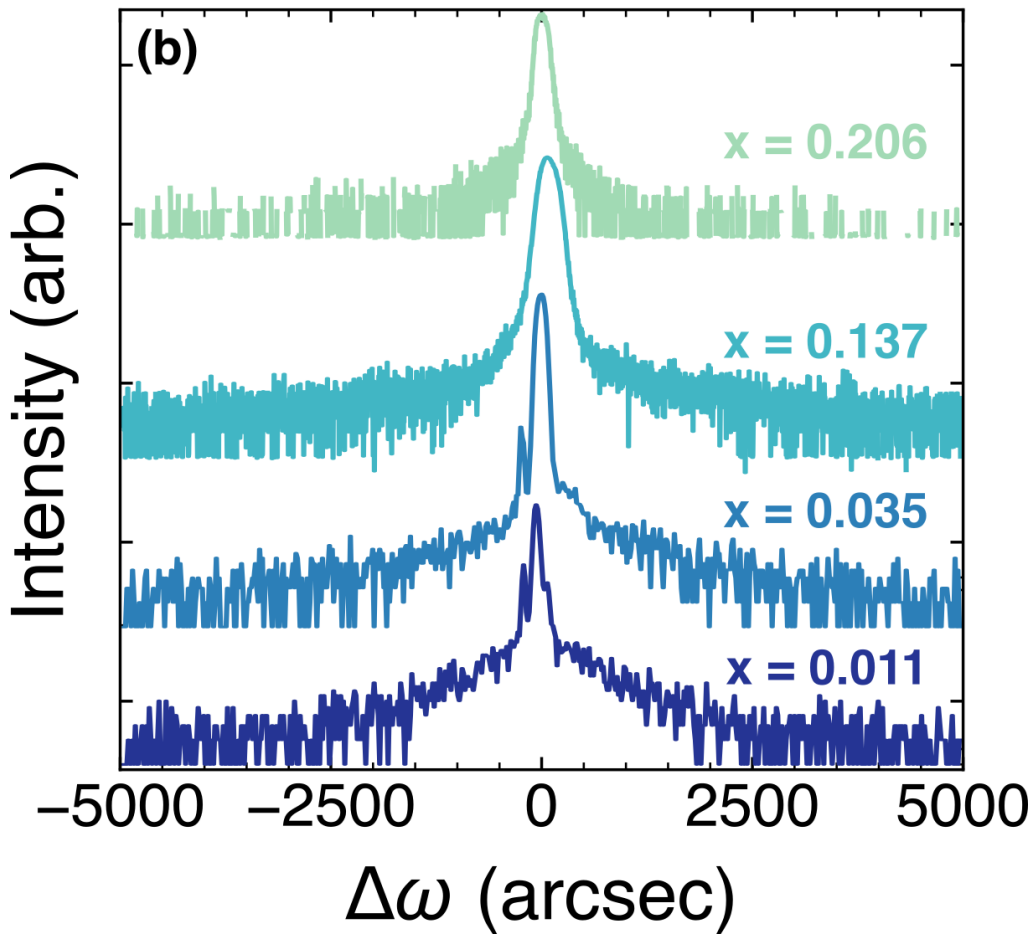

Figure 1. **X-ray diffraction analysis of $(Al,Ga)_2O_3$ epilayers shows high crystalline quality.** (a) symmetric XRD scans of the monoclinic (020) reflection show strong thickness oscillations indicating out-of-plane coherence in the epilayers. (b) Rocking curves of (020) epilayer peak are narrow, demonstrating a low degree of mosaic tilt.

Figure 1a shows symmetric XRD of the films, demonstrating high crystalline uniformity and smooth surfaces through sharp peaks with thickness fringes. The fringes give thicknesses ranging from 82 – 130 nm (see Table S1. in Supporting Information for details), and growth rates from 3.5 – 4.3 nm/min. Survey scans (Figure S1.) demonstrate only β-phase monoclinic peaks. The aluminum content was obtained by the position of the XRD peak assuming a strained epilayer.[33]

The use of indium as a catalyst for growth was critical to achieving high-quality films at increased Al-composition. Previous studies have demonstrated that indium can increase growth rate and accessable growth temperatures without incorporating into the film.[32] In this process, $In_2O_3$ is more favorable to form than $Ga_2O_3$, bypassing the formation of volatile $Ga_2O$ that occurs during growth without indium. A cation exchange occurs in the presence of Ga, with $In_2O_3$ converting to $Ga_2O_3$ and excess In desorbing from the surface. Thus, In acts to suppress $Ga_2O$ desorption, and also as a surfactant to modify the surface energy to promote smooth growth. This indium-catalyzed growth process enables $(Al_xGa_{1-x})_2O_3$ growth at higher substrate temperatures by increasing adatom diffusion, resulting in a lower extended defect density due to increased adatom diffusion for x up to 0.2 in $(Al_xGa_{1-x})_2O_3$) on (010)-oriented substrates. Previous reports of $(Al_{0.2}Ga_{0.8})_2O_3$ grown by MBE without indium are poor quality with secondary phases.[34] The coherence and low extended defect density of the $(Al_xGa_{1-x})_2O_3$ epilayers is corroborated by narrow (020) rocking curves as shown in Figure 1b, with full-widths at half maximum (FWHM) < 250". In the x= 0.011 and x = 0.035 rocking curves, the slits could not be narrowed enough to fully exclude the substrate peak directly to the left of the epilayer peak. Rocking curves were also obtained around the (420) asymmetric reflection to measure twist mosaicity, and the FWHM are all < 350" (largest for the x = 0.206 film) indicating excellent crystal quality. Individual tabulated (020) and (420) rocking curve widths are given in the Table S1.

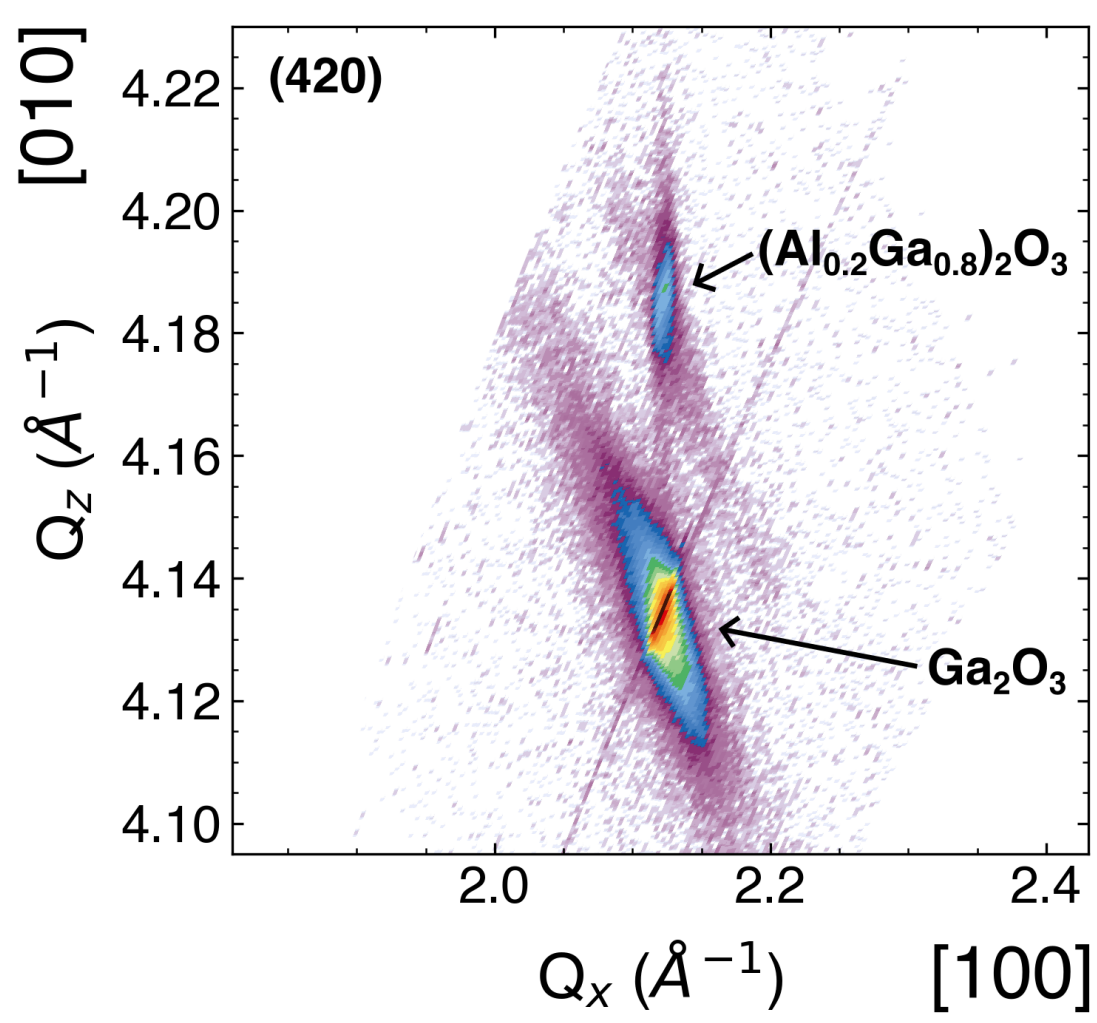

Figure 2. **$(Al,Ga)_2O_3$ epilayers are strained to the β-$Ga_2O_3$ substrate.** Reciprocal space map taken around the (420) reflection of the highest composition is explored for $(Al_{0.2}Ga_{0.8})_2O_3$. Strain along the [100] direction is demonstrated by identical in-plane $Q_x$ vectors. The artifact through the substrate is the analyzer streak.

A reciprocal space map around the (420) reflection is shown in Fig. 2 to confirm the pseudomorphic nature of the film with respect to the substrate. The artifact running from bottom-left to top-right is analyzer streak that arises from finite beam divergence in the ω-axis. The vertical elongation of the $(Al_{0.2}Ga_{0.8})_2O_3$ reflection is the crystal truncation rod arising from finite thickness broadening. ω-axis (rocking) smearing is observed in the substrate, indicating substrate defects with rotational mosaicity. This broadening is reflected in the epilayer at very low intensity, indicating that these defects are not strongly propogated into the film. Vertical alignment along the in-plane $Q_x$ reciprocal lattice vector indicates the films are pseudomorphic along the [100] direction. For the highest aluminum content at x = 0.206, the film is 0.7% strained along [100] and 0.6% strained along [001]. Pseudomorphic growth is important for alloy crystal quality to avoid misfit dislocations which can intensify phonon scattering. Overall, the XRD analysis of the $(Al_xGa_{1-x})_2O_3$ epilayers indicate highly uniform crystallinity with minimal dislocation density or mosaicity.

Thermal properties of $(Al_xGa_{1-x})_2O_3$ epilayer on $\beta$-$Ga_2O_3$ (010) substrate were measured using the time-domain thermoreflectance (TDTR) technique.[35,36] Details of the set up can be found in Supporting Information. Prior to the experiment, a thin Al transducer of 80 nm is deposited with its thermal conductivity estimated by four-point probe, and its thickness confirmed through picosecond acoustics.

To extract the thermal properties of Al/$(Al_xGa_{1-x})_2O_3$/$\beta$-$Ga_2O_3$, the experimental ratio signal is fit to a cylindrically symmetric heat equation model[37] (as shown in Figure 3) that accounts for individual layers' thickness, thermal conductivity and volumetric heat capacity along with thermal boundary conductance across the interfaces.

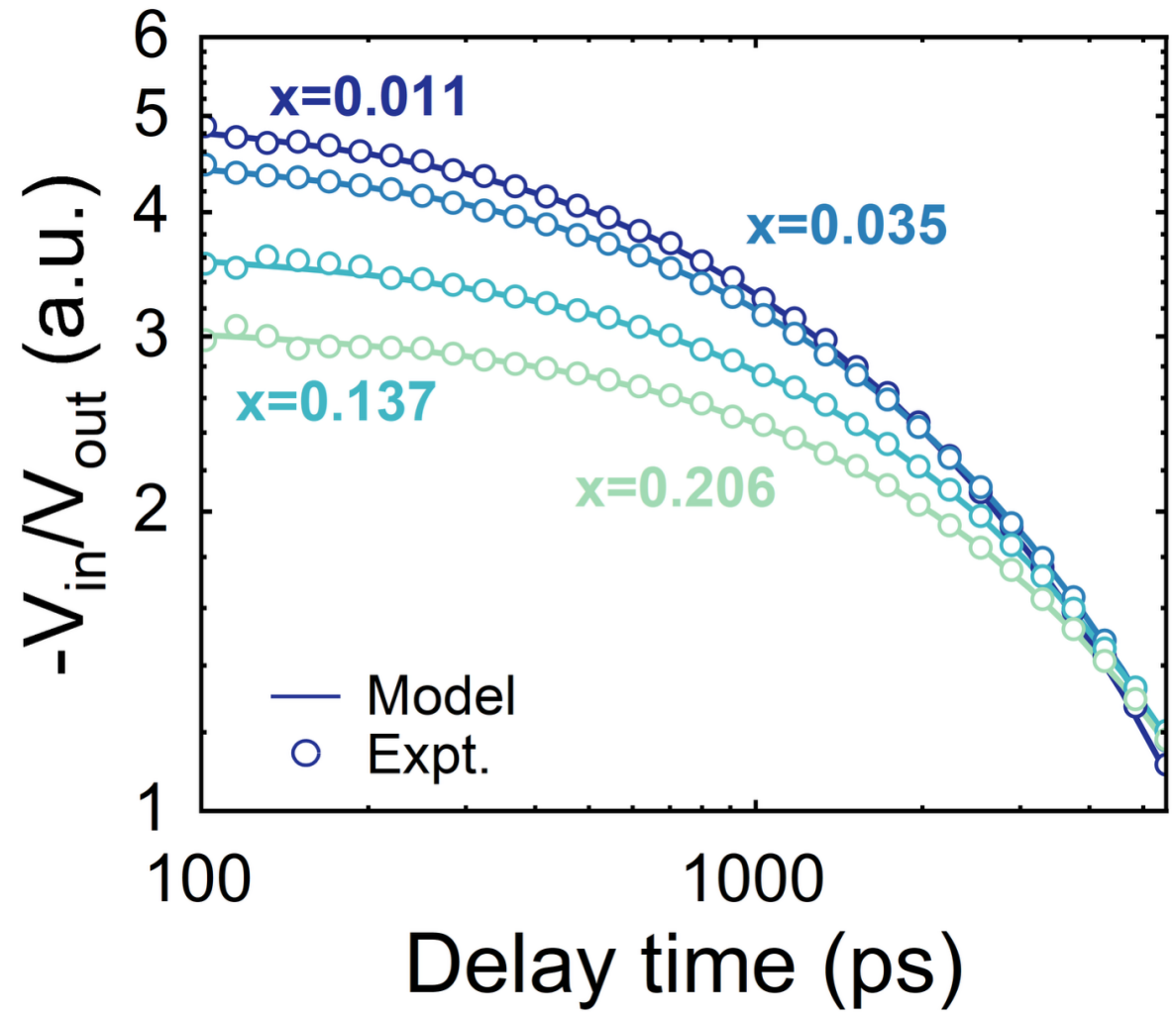

Figure 3. **Signal ratio "$-V_{in}/V_{out}$" as a function of time delay for Al/$(Al_xGa_{1-x})_2O_3$/$\beta$-$Ga_2O_3$ heterostructure at a modulation frequency of 8.4 MHz**. Data are shown for varying Al compositions, x = 0.011 (black), 0.035 (red), 0.137 (blue), and 0.206 (green). Open circles represent experimental data and solid lines represent model fits. A reduction in $-V_{in}/V_{out}$ magnitude is observed with increase in Al composition from 0.011 to 0.206 in $(Al_xGa_{1-x})_2O_3$ film.

A literature value of 2.9 MJ $m^{-3}$ $K^{-1}$ was assumed for the volumetric heat capacity for both $\beta$-$Ga_2O_3$ and $(Al_xGa_{1-x})_2O_3$.[38] The cross-plane thermal conductivity of the (010) $\beta$-$Ga_2O_3$ substrate was measured from a control sample to be 22 $\pm$ 2 W $m^{-1}$ $K^{-1}$, which is slightly lower than literature value reported for both doped and undoped (010) $\beta$-$Ga_2O_3$[13]. The measurement is minimally sensitive to the in-plane thermal conductivity, which is assumed as 10.9 W $m^{-1}$ $K^{-1}$.[39]

The fitted parameters include the cross-plane thermal conductivity $(\kappa)$ of the $(Al_xGa_{1-x})_2O_3$ film and thermal boundary conductance $(G_1)$ across Al/$(Al_xGa_{1-x})_2O_3$ interface. The thermal boundary conductance $(G_2)$ of the $(Al_xGa_{1-x})_2O_3$/$\beta$-$Ga_2O_3$ interface is assumed to be negligibly high due to lack of sensitivity as shown in Figure S2 and S3. To fit for $\kappa$ and $G_1$, we perform perform analysis at 8.4 MHz based on the sensitivity calculations (Figure S2 and S3). The sensitivity to the films thermal conductivity varies with respect to the composition and the uncertainty varies accordingly, as discussed below.

Figure 4a shows the nominal fitted values for the cross-plane thermal conductivity $(\kappa)$ of the (010) oriented $(Al_xGa_{1-x})_2O_3$ epilayer films as a function of Al concentration x. Measurement uncertainties were determined using Monte Carlo simulations in a similar manner as Ref [40]. We find the Monte Carlo results for the thermal conductivity yield a non-symmetric distribution. Therefore, we define the uncertainty range from the 95% single-sided lower and upper bounds. That is, 95% of the simulated fit results are above or below these bounds, respectively. For x = 0.011 and 0.035, the calculated upper limit in the fit uncertainty is unbounded due to limitations in measurement sensitivity, For example, for x < 0.05, sensitivity calculations (Figure S2.) indicate that the high magnitude ratio signal (Figure 4) across the time delay could be due to the thermal resistance of the thin film being dominated by the substrate, even at a modulation frequency of 8.4 MHz.

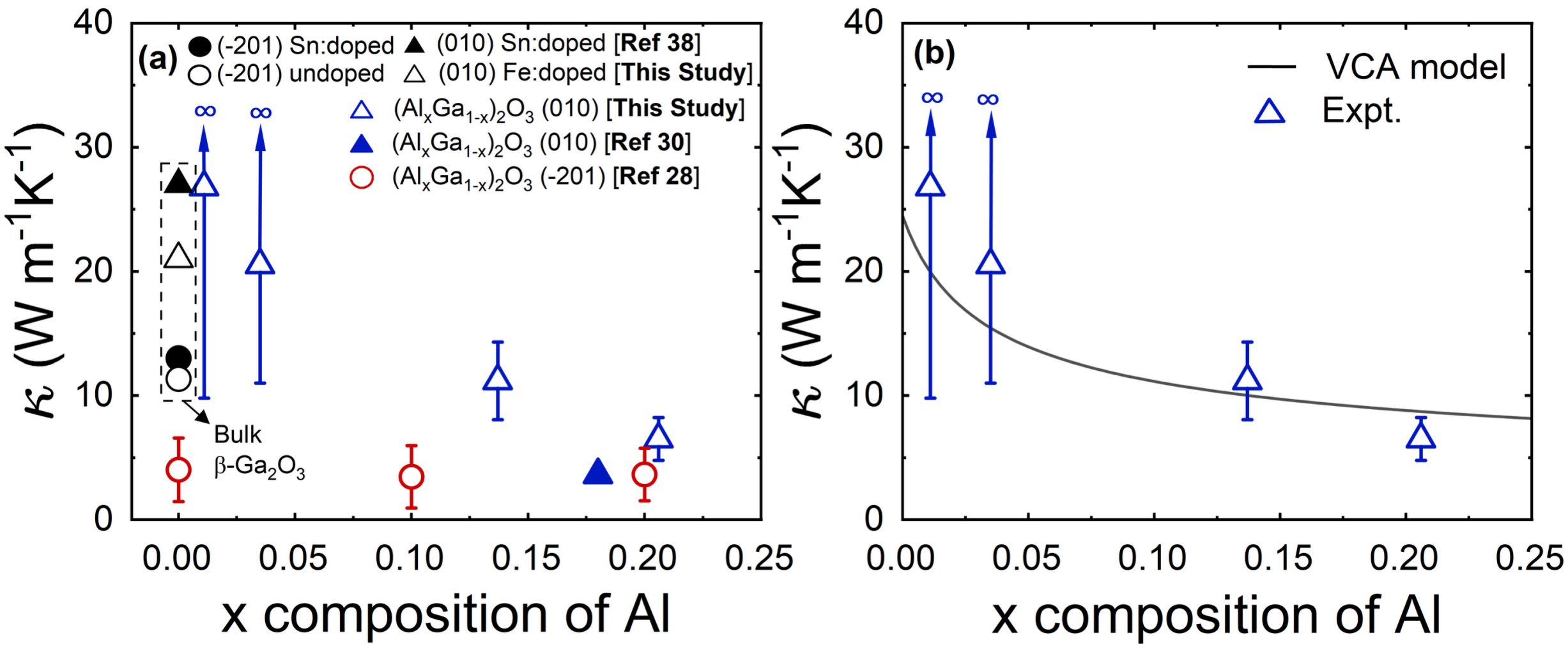

Figure 4. **Cross-plane thermal conductivity ($\kappa$) of $(Al_xGa_{1-x})_2O_3$ (010) thin film measured as a function of Al concentration** (a) compared to previously reported (-201) and (010) oriented $(Al_xGa_{1-x})_2O_3$ thin films from ref [28] and [30], respectively; In addition, bulk $\beta$-$Ga_2O_3$ for (010) and (-201) is included from current study and from ref [38] to show size and alloy scattering effects. (b) Thermal conductivity compared to the virtual crystal approximation (VCA) model validates alloy scattering across Al composition. The uncertainity measurements for this study was performed using monte Carlo analysis. The blue triangles represent nominal fitted value; the error bars represent 95% monte carlo bounds. For low x the upper uncertainty limit is unbounded due to limited contrast between substrate and epilayer.

To interpret these findings, the thermal conductivity measured in this study was compared with a previously reported $\kappa$ value for (010) oriented $(Al_{0.18}Ga_{0.82})_2O_3$ film of thickness 330 nm (Figure 4a).[30] The thermal conductivity from ref [30] is roughly two times lower than the value we report for $(Al_{0.137}Ga_{0.863})_2O_3$ and $(Al_{0.206}Ga_{0.794})_2O_3$ films and the difference can be understood by analyzing the XRD. The previous report exhibits phase-separated $(Al_{0.18}Ga_{0.82})_2O_3$ showing a (440) γ-phase (spinel) peak alongside the (020) monoclinic substrate and epilayer peaks (supplementary XRD ref [30]). Our films at comparable concentrations show no phase separation at the detection limit of XRD (Supporting Information, Figure S1.) This suggests that the domains and dislocations associated with γ-phase inclusions in $(Al_xGa_{1-x})_2O_3$ significantly degrade overall thermal transport properties.

To understand the role of crystallographic orientation, we compare our (010) oriented $(Al_xGa_{1-x})_2O_3$ thermal conductivity against previously reported (-201) oriented films.[28] As illustrated in Figure 4a, the thermal conductivity of (-201) films remain nearly composition-independent in contrast to the strong composition dependence observed in this work. This discrepancy likely indicates that alloy scattering dominates heat diffusion in the (010) orientation as Al fraction increases, whereas this mechanism plays a far less pronounced role in the previously reported (-201) oriented films.

To test the hypothesis that the compositional trend of thermal conductivity ($\kappa$ vs x) in (010) oriented $(Al_xGa_{1-x})_2O_3$ is governed by phonon-scattering due to atomic mass differences, we compare our experimental results to theoretical lattice thermal conductivity predictions using a Virtual Crystal Approximation (VCA) framework.[41,42] Under this Debye Callaway-type model within the relaxation-time approximation, lattice thermal conductivity is expressed as:

$$\kappa(T,x) = \frac{k_\mathrm{B}}{2\pi^2 v}\left(\frac{k_\mathrm{B}T}{\hbar}\right)^3 \int_0^{\Theta_\mathrm{D}/T} \tau\,(y,T,x)\frac{y^4 e^y}{(e^y-1)^2}\,\mathrm{d}y, \quad \omega = \frac{y k_\mathrm{B} T}{\hbar} \qquad (1)$$

Here, $k_\mathrm{B}$ is the Boltzmann constant, $T$ is temperature, $\hbar$ is the reduced Planck constant, $v$ is an effective acoustic velocity, $\Theta_\mathrm{D}$ is the Debye temperature (738 K),[39] and $x$ is the composition. Independent scattering processes were combined using Matthiessen's rule,

$$\tau^{-1}(\omega,T,x) = \tau_\mathrm{U}^{-1} + \tau_\mathrm{bulk}^{-1} + \tau_\mathrm{alloy}^{-1}. \qquad (2)$$

with $\tau_\mathrm{U}^{-1} = B_\mathrm{U} T\omega^2 \exp\left(-\frac{C_\mathrm{U}}{T}\right)$ and $\tau_\mathrm{bulk}^{-1} = A_\mathrm{bulk}\omega^4$ being the Umkapp and bulk defect scattering rates in $\beta$-$Ga_2O_3$, respectively. The Umklapp and bulk scattering terms were determined from the temperature- dependent (010) thermal conductivity of bulk $\beta$-$Ga_2O_3$ reported by Guo *et al.*[39] A nonlinear least-squares fit of the data yields $B_\mathrm{U} = 2.214 \times 10^{-18}\ \mathrm{s\,K^{-1}}$, $C_\mathrm{U} = 560.1\ \mathrm{K}$, and $A_\mathrm{bulk} = 1.084 \times 10^{-42}\ \mathrm{s^3}$.

Random substitution of Al on the Ga sublattice was represented by a Klemens-type Rayleigh mass-disorder term,[43,44]

$$\tau_\mathrm{alloy}^{-1} = P\,\Gamma_M(x)\omega^4, \qquad P = \frac{\mathrm{V}}{4\pi v^3}. \qquad (3)$$

where the binary cation mass-variance parameter is

$$\Gamma_M(x) = x(1-x)\left[\frac{\Delta M}{\overline{M}_c(x)}\right]^2, \qquad \overline{M}_c(x) = (1-x)M_\mathrm{Ga} + xM_\mathrm{Al}, \quad \Delta M = M_\mathrm{Ga} - M_\mathrm{Al}. \qquad (4)$$

The standard atomic masses $M_\mathrm{Ga} = 69.723\ \mathrm{u}$ and $M_\mathrm{Al} = 26.9815385\ \mathrm{u}$ were used from Ref [40]. We apply an effective scatterer volume of $\mathrm{V} = 104.425\ \mathrm{Å}^3$, corresponding to the primitive-cell volume of $\beta$-$Ga_2O_3$ [25].

For the sound speeds, the $[010]$ longitudinal acoustic velocity, $v_L$, was assumed to be independent of Al concentration, whereas the two transverse velocities were treated as varying with the Al fraction based up on the trends in Ref [45]:

$$v_L = 7800\,\frac{m}{s},$$

$$v_{T1}(x) = 4100 + \frac{x}{0.2}(6200 - 4100)\frac{m}{s},$$

$$v_{T2}(x) = 3000 + \frac{x}{0.2}(5500 - 3000)\frac{m}{s},$$

Where $x$ is expressed as a fractional cation concentration. An effective velocity was then obtained with an inverse square average using a similar approach as Wang and Mingo from Ref [42]

$$v(x) = \left(\frac{1}{3}\left(v_L^{-2} + v_{T1}(x)^{-2} + v_{T2}(x)^{-2}\right)\right)^{-1/2}.$$

With these values, we generate the aluminum concentration-dependent thermal conductivity model shown in Figure 4b. Despite the simplicity of the model, we find reasonable agreement with the experimentally obtained thermal conductivity, indicating that the thermal conductivity changes of our (010) oriented $(Al_xGa_{1-x})_2O_3$ are driven by phonon scattering due to the mass differences between the Ga and Al.

The thermal boundary conductance across the $Al/(Al_xGa_{1-x})_2O_3$ interface ($G$) as a function of Al composition (x) is show in Figure 5. There is a monotonic decline in thermal boundary conductance as Al content is increased. This suggests that the increased Al content modifies the phonon transmission, reducing the heat flow from Al to $(Al_xGa_{1-x})_2O_3$.

The physical mechanism governing phonon thermal boundary conductance across metal/non-metal interfaces has been reviewed extensively in the literature over the past few decades.[46,47] In a simplified, qualitative picture of elastic interfacial transmission, phonons in the Al film transmit energy exclusively to phonons of identical frequencies in $(Al_xGa_{1-x})_2O_3$. Within this elastic scattering limit, phonon modes in $(Al_xGa_{1-x})_2O_3$ with frequencies exceeding the maximum cutoff frequency of Al do not contribute to heat transport across the interface. Consequently, under the diffuse mismatch model (DMM), variations in the phonon density of states (DOS) mismatch between Al and $(Al_xGa_{1-x})_2O_3$ directly dictate changes in the thermal boundary conductance.

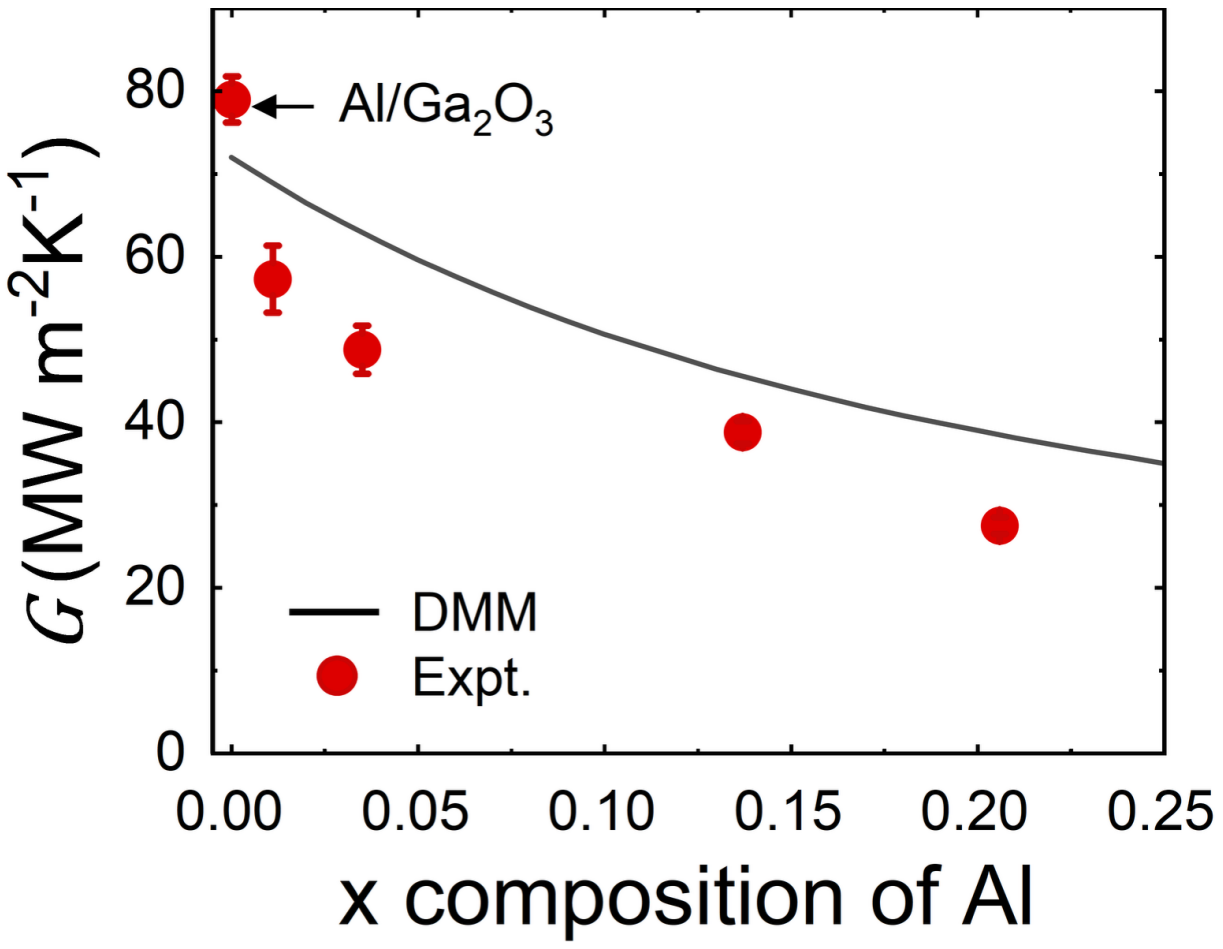

Figure 5. **Thermal boundary conductance (*G*) measured across $Al/(Al_xGa_{1-x})_2O_3$ interface as a function of Al concentration.** The solid circles are our measured values and the lien is the preciction from a DMM calculation assuming a Debye solid for the $(Al_xGa_{1-x})_2O_3$ and a sine-type dispersion for the Al. Our results suggest that the acoustic phonon mode stiffening in the $(Al_xGa_{1-x})_2O_3$, which spectrally overalap with the phonon modes in the Al, gives leads to reduced transmission of phonon energies from the Al to $(Al_xGa_{1-x})_2O_3$ and concomitant reduction in thermal boundary conductance. The uncertainity for these measurements was performed using Monte Carlo analysis described in the text and Supporting Information.

To test this hypothesis, we model the thermal boundary conductance with a Diffuse Mismatch Model (DMM). We assume a Debye approximation for the three lowest acoustic modes in the $(Al_xGa_{1-x})_2O_3$ and a sine-type approximation for the acoustic modes in the Al transducer, analogous to our virtual crystal model used to predict $k$ versus $x$. Key to our model is that the transverse acoustic modes of $(Al_xGa_{1-x})_2O_3$ are predicted to stiffen with increased x,[45] and thus the acoustic mode phonon velocities increase with increased Al composition, as mentioned previously in our discussion of our virtual crystal model.

The DMM predictions for thermal boundary conductance as a function of x across the $Al/(Al_xGa_{1-x})_2O_3$ interface are shown in Fig. 5, along with our experimental measurements. Our model reflects the general trends observed in our experimental data of reduced thermal boundary conductance versus $x$. While we also note that our model achieves acceptable quantitative agreement, we caution that our model for the phonon spectra of $(Al_xGa_{1-x})_2O_3$ is based on Debye solid of only the three acoustic modes, so we do not consider the optical phonon modes in $(Al_xGa_{1-x})_2O_3$ within the bandwidth of the phonon spectra in Al that could be contributing to the interfacial conductance.[48] While this could impact the DMM-predicted values for thermal boundary conductance, we do not expect this trend versus $x$ to change, as the primary changes in heat flux versus Al composition in the $(Al_xGa_{1-x})_2O_3$ arises in the TA modes. Thus, even if optical phonon modes in the $(Al_xGa_{1-x})_2O_3$ are stiffening, the small group velocities of these modes will not lead to appreciable changes in the DMM-predicted trends of conductance versus $x$. Our results suggest that the acoustic phonon mode stiffening in $(Al_xGa_{1-x})_2O_3$, which spectrally overlaps with the phonon modes in the Al, gives rise to reduced transmission of phonons from the Al to $(Al_xGa_{1-x})_2O_3$ and concomitant reduction in thermal boundary conductance.

In conclusion, the cross-plane thermal conductivity ($\kappa$) of (010) oriented $(Al_xGa_{1-x})_2O_3$ thin film on $\beta$-$Ga_2O_3$ substrate was measured using TDTR for x composition ranging 0.011 to 0.206. A reduction in $\kappa$ was observed due to increased alloy scattering with respect to the Al composition, which was validated using a VCA model. We further examined the thermal boundary conductance across $Al/(Al_xGa_{1-x})_2O_3$ interface which was found to reduce with increase in x due to a mismatch in acoustic phonon density of states that changes with x. The overall thermal conductivity of the $(Al_xGa_{1-x})_2O_3$ alloys were >2X than those previously reported. This increase in $\kappa$ is attributed to indium-catalyzed MBE growth which eliminates phase separation and reduces extended defect density for the compositions studied. These findings are an important step towards the design, fabrication, and optimization of industrially-relevant $(Al_xGa_{1-x})_2O_3/Ga_2O_3$ devices for high-power and high-frequency applications.

**Acknowledgements**

The work was primarily supported as part of A Center for Power Electronics Materials and Manufacturing Exploration (APEX), an Energy Frontier Research Center funded by the U.S. Department of Energy, Office of Science, Basic Energy Sciences. Thermal conductivity and X-ray diffraction experiments were supported as part of APEX. Thin film synthesis was supported by the Laboratory Directed Research and Development (LDRD) Program at the National Laboratory of the Rockies. This work was authored in part by the National Laboratory of the Rockies for the U.S. Department of Energy (DOE), operated under Contract No. DE-AC36-08GO28308. The views expressed in the article do not necessarily represent the views of the DOE or the U.S. Government.

# Supporting Information

## Enhanced thermal conductivity of (010) $(Al_xGa_{1-x})_2O_3$ epitaxial films utilizing indium-catalyzed molecular beam epitaxy

Shivashree Gowda[1], Stephen Schaefer[2], Ethan A. Scott[1], Samreen Khan[1], Patrick E. Hopkins[1,3,4,a)], M. Brooks Tellekamp[2,a)]

[1]Department of Mechanical and Aerospace Engineering, University of Virginia, Charlottesville, VA 22904, USA

[2]National Laboratory of the Rockies, Golden, CO 80401, USA

[3]Department of Material Science and Engineering, University of Virginia, Charlottesville, VA 22904, USA

[4]Department of physics, University of Virginia, Charlottesville, VA 22904, USA

[a)]**Author to whom correspondence should be addressed**:
brooks.tellekamp@nlr.gov, peh4v@virginia.edu

### MBE Growth and characterization

### S1. $(Al_xGa_{1-x})_2O_3$ growth conditions

Films were grown in a Riber Compact 21T MBE using standard effusion cells to source 6N5 Al, 7N Ga, and 7N In. Oxygen was sourced using a 13.56 MHz radio-frequency oxygen plasma from a quartz bulb. Fluxes were measured using a retractable ion gauge beam-flux monitor. The substrates were cleaned in acetone, methanol, and isopropanol, followed by a deionized water rinse and an organic etch using $H_2SO_4$:$H_2O_2$ 4:1 heated at 140 °C for 10 minutes. Samples were outgassed at 150 °C overnight before loading into the growth chamber with a background pressure <1E-10 torr. The samples were outgassed at 800 °C for 10 minutes, and then etched using 4E-7 torr beam equivalent pressure (BEP) gallium corresponding to an etch rate of approximately 6 nm/min. After etching the the substrates were exposed to active oxygen for 15 minutes. Then, the substrate temperature was adjusted and sources were ramped to growth conditions. The Ga flux was held constant at 1.5E-7 torr BEP, the In flux at 2E-7 torr BEP, and the Al flux was varied from 1E-9 torr to 1.8 E-8 torr BEP to control the Al concentration in the alloy. All films were grown for 30 minutes with 3 SCCM flow of $O_2$ with 250W RF power, except for the $(Al_{0.2}Ga_{0.8})_2O_3$ sample which was grown at 2 SCCM $O_2$ for 23 minutes. The substrate temperature ranged from 700 – 750 °C, measured using a calibrated UV band-edge pyrometer (k-Space Associates).

*Table S1. Measured aluminum content, thickness, and rocking curve widths for on-axis (020) and off-axis (420) reflections.*

| Al-content x | Thickness | (020) FWHM | (420) FWHM |
|---|---|---|---|
| 0.011 | 112 nm | 71” | † |
| 0.035 | 116 nm | 122” | 212” |
| 0.137 | 130 nm | 248” | 213” |
| 0.206 | 82 nm | 181” | 350” |

† - cannot definitively separate epilayer and substrate peak

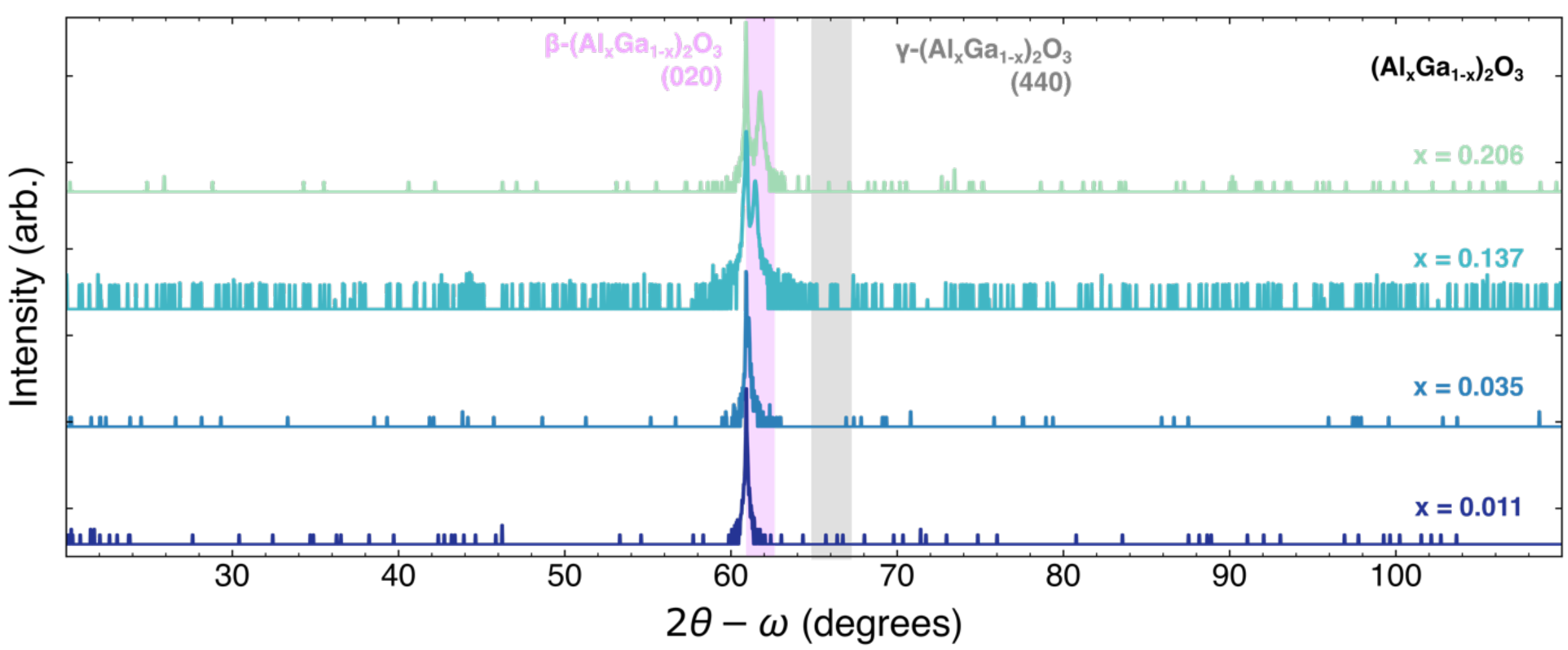

*Figure S1. Wide angle 2θ-ω survey scans demonstrating single-phase epilayers up to x=0.206. Phase separation typically occurs by formation of (440)-oriented γ-phase spinel. Both β-phase (purple) and γ-phase (grey) peak ranges are highlighted in the figure for the range of possible aluminum compositions. γ phase is not observed.*

## <u>Thermal Characterization</u>

### S2. Time-domain thermoreflectance (TDTR) setup description

In TDTR, the output of a femtosecond laser is spectrally separated into distinct pump and probe beams. The pump beam is modulated using an electro-optic modulator. The probe beam is optically delayed with respect to the pump. Both the pump and probe beam are coaxially focused on to the sample surface with a $1/e^2$ beam diameter of 12 µm. The modulated pump periodically heats the sample, and the probe measures the resulting temperature-induced change in surface reflectance. The reflected probe beam is detected by a lock-in amplifier referenced to the pump modulation frequency to extract in-phase ($V_{in}$) and out-of-phase ($V_{out}$) voltage components. The $-V_{in}/V_{out}$ ratio signal is fit to a heat diffusion equation-based model to obtain the thermal properties of interest.

*Tabel S2. Material properties assumed for the sensitivity models. The thermal boundary conductance ($G$) between ($Al_xGa_{1-x})_2O_3$ film and substrate was assumed to be negligibly high. The film thermal conductivity and the interface conductance between the transducer are fit parameters. Here $k_z$ and $k_r$ refers to cross-plane and in-plane thermal conductivity, respectively; $C_v$ refers to volumetric heat capacity; h refers to thickness.*

| Layer | Material | $k_z$ (W $m^{-1}K^{-1}$) | $k_r$ (W $m^{-1}K^{-1}$) | $C_v$ (MJ $m^{-3}K^{-1}$) | h (nm) | $G$ (MW $m^{-2}K^{-1}$) |
|---|---|---|---|---|---|---|
| 1 | Al | 163 | 163 | 2.42 | 80 | fit |
| 2 | $(Al_xGa_{1-x})_2O_3$ film | fit | $k_z$ | 2.9 | (Refer to Table S1) | 1E20 |
| 3 | Substrate | 21.8 | 10.9 | 2.9 | 1E09 | - |

## S3. Sensitivity analysis

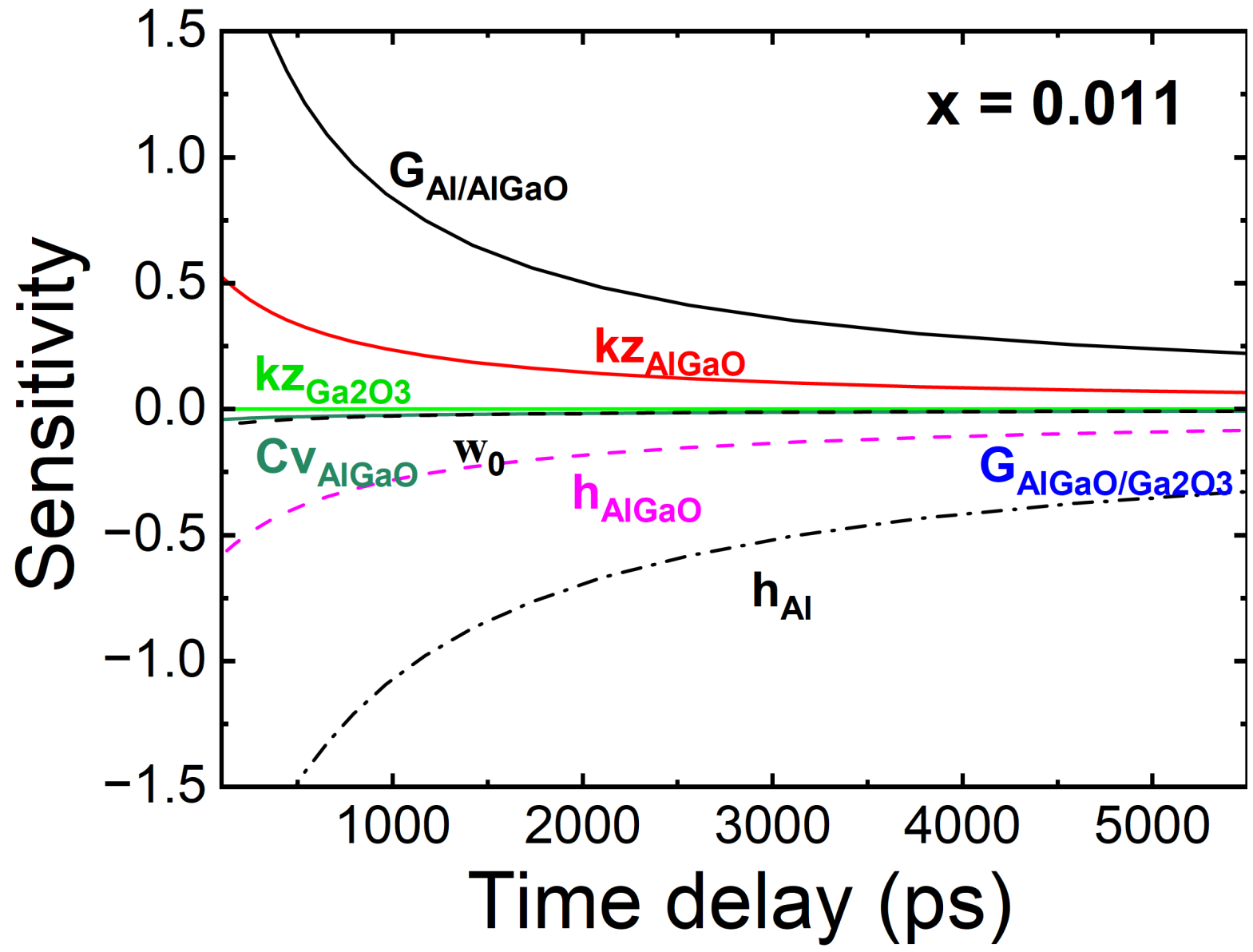

*Figure S2. Sensitivity plots for thermal measurement of $Al/(Al_xGa_{1-x})_2O_3/\beta$-$Ga_2O_3$ for x composition of 0.011 at 8.4 MHz. Here, kz is cross-plane thermal conductivity; Cv is volumetric heat capacity; G is thermal boundary conductance; h is the thickness of the film (AlGaO) and the Al transducer (Al); $\omega_0$ is convoluted pump and probe spot size. The sensitivity curves assume the material properties from Table S2.*

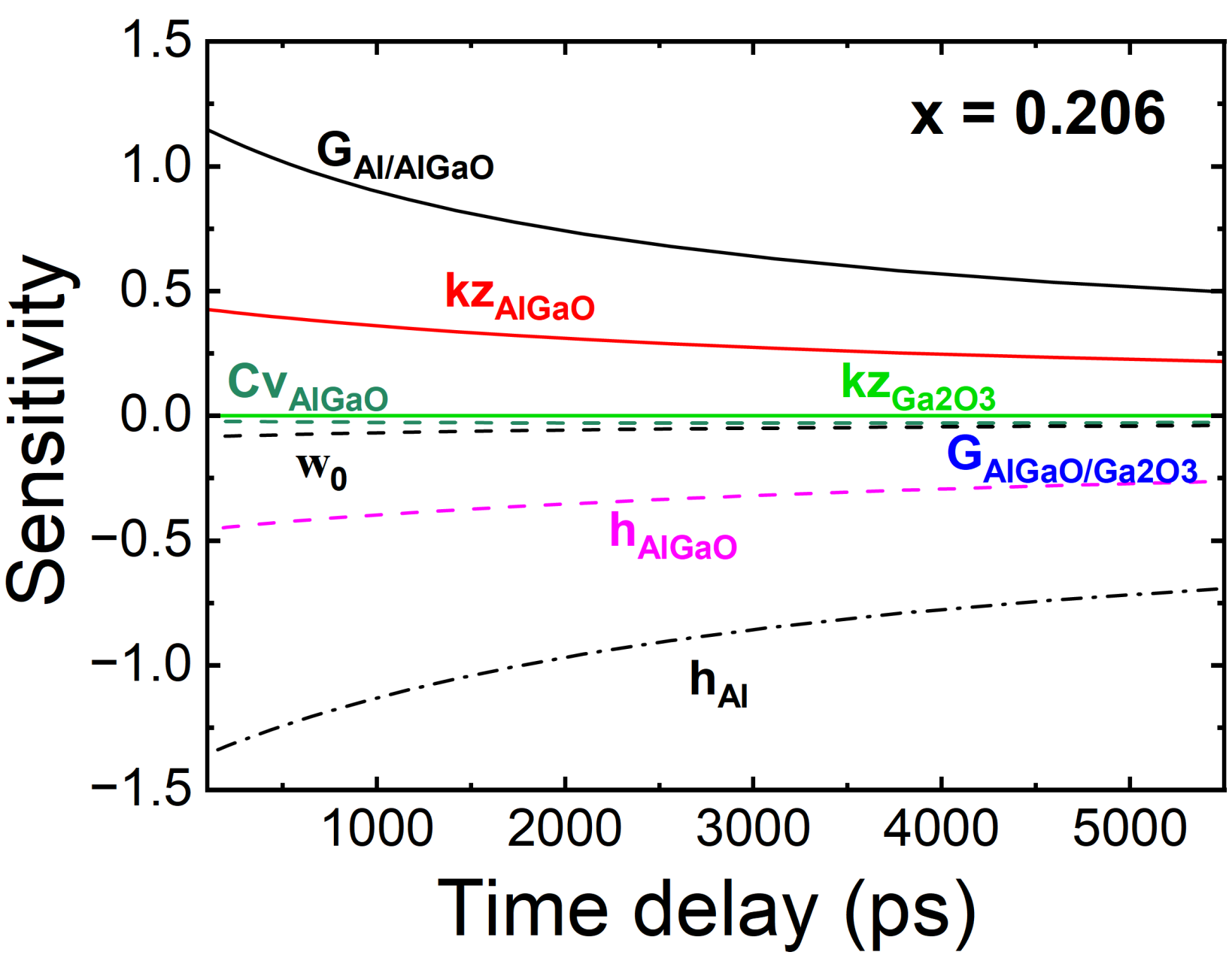

*Figure S3. Sensitivity plots for thermal measurement of Al/$(Al_xGa_{1-x})_2O_3$/β-$Ga_2O_3$ for x composition of 0.206 at 8.4 MHz. The sensitivity curves assume the material properties from Table S2.*

## S4. Monte Carlo Analysis

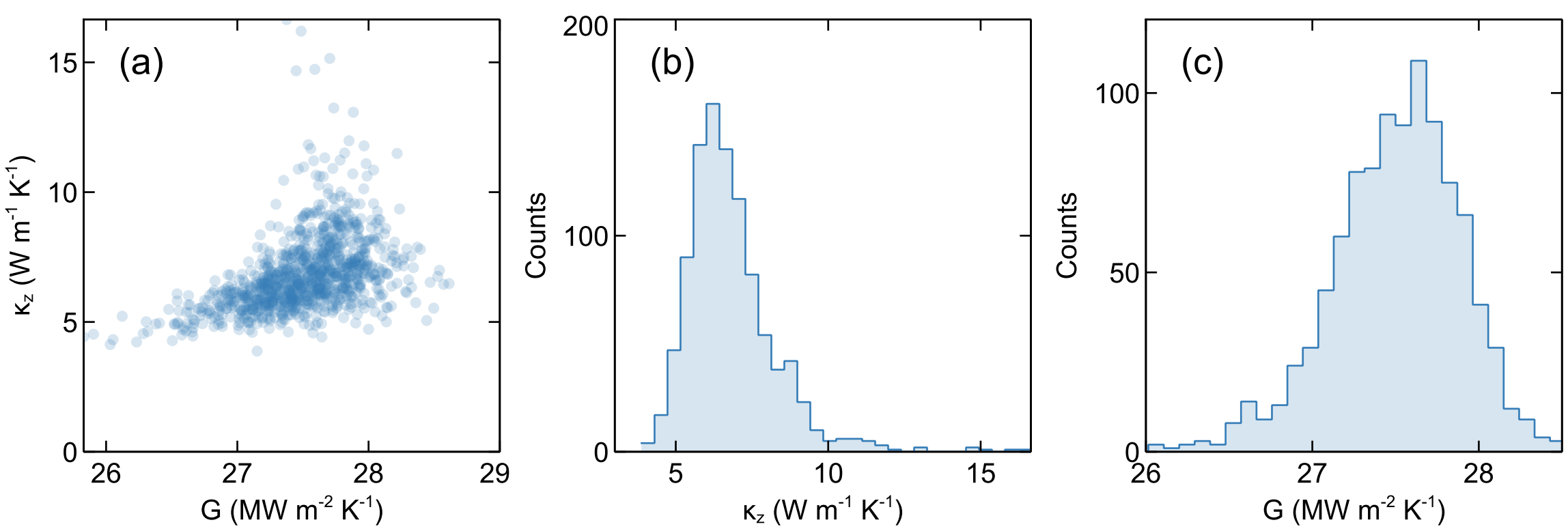

Figure S4. Example Monte Carlo analysis of the x = 0.206 composition for N = 1000 iterations. The simulation accounts for variation in the measured Al thermal conductivity (±10 W m$^{-1}$K$^{-1}$) and thickness (±3 nm) as well as the uncertainty in the measured substrate cross-plane thermal conductivity (±1.7 W m$^{-1}$K$^{-1}$). (a) shows the distribution of fitted thermal conductivity values and associated thermal boundary conductances for each iteration. (b) and (c) show histograms for the thermal conductivity and thermal boundary conductance distributions, respectively. Given the non-symmetric distributions, we define the uncertainty range as the 95% one-sided lower and upper bounds. That is, 95% of the fitted results lie above or below these bounds, respectively.